\documentclass{ledger}

\usepackage{xcolor}
\definecolor{ngreen}{rgb}{0,0.5,0}

\newcommand{\thefirstpagenum}[0]{1}

\newcommand{\thedoi}[0]{0}

\newcommand{\beq}{\begin{equation}}
\newcommand{\eeq}{\end{equation}}
\newcommand{\bqa}{\begin{eqnarray}}
\newcommand{\eqa}{\end{eqnarray}}

\newcommand{\blk}{\color{black}}

\hypersetup{pdfauthor={A. J. Bennet}, pdftitle={Energy efficient mining on a quantum--enabled blockchain using light}}

\title{Energy efficient mining on a quantum--enabled blockchain using light}
\author{Adam~J.~Bennet\thanks{A.~J.~Bennet is an independent researcher in Melbourne, Australia.}
\and Shakib~Daryanoosh\thanks{S.~Daryanoosh is a post--doctoral researcher at Macquarie University, Department of Physics and Astronomy, and the Australian Research Council Centre of Excellence for Engineered Quantum Systems, Sydney, Australia.}}

\pretitle{
  \centering \selectfont  \par
  \fontsize{20pt}{24pt}\selectfont} 

\begin{document}

\maketitle

\thispagestyle{pagefirst}

\begin{abstract}
We outline a quantum-enabled blockchain architecture based on a consortium of quantum servers. The network is hybridised, utilising digital systems for sharing and processing classical information combined with a fibre--optic infrastructure and quantum devices for transmitting and processing quantum information. We deliver an energy efficient interactive mining protocol enacted between clients and servers which uses quantum information encoded in light and removes the need for trust in network infrastructure. Instead, clients on the network need only trust the transparent network code, and that their devices adhere to the rules of quantum physics. To demonstrate the energy efficiency of the mining protocol, we elaborate upon the results of two previous experiments (one performed over 1km of optical fibre) as applied to this work. Finally, we address some key vulnerabilities, explore open questions, and observe forward--compatibility with the quantum internet and quantum computing technologies.

\end{abstract}

\section{Introduction}
Blockchain technology has shown potential for transforming traditional industry with its key characteristics of decentralization, transparency, persistency, and auditability. Despite the great potential of blockchain, it faces numerous challenges, which limit the wide use of blockchain technology. For example, the energy consumption of the Bitcoin blockchain\endnote{Nakamoto, S. ``Bitcoin: A Peer-to-Peer Electronic Cash System.'' \textit{Bitcoin.org} (2009) (accessed August 2018) \url{http://bitcoin.org/bitcoin.pdf}.}\endnote{Antonopoulos, A. M. \textit{Mastering Bitcoin: Unlocking Digital Cryptocurrencies.} O'Reilly Media, Inc. (2015)}\endnote{Extance, A. ``The future of cryptocurrencies: Bitcoin and beyond.'' \textit{Nature} \textbf{526} 21--23 (2015).}  has been predicted to eclipse total global energy consumption by 2020.\endnote{de Vries, A. ``Bitcoin's growing energy problem''. \textit{Joule}, \textbf{2.5}, 801--805 (2018).} Blockchain technology also faces issues with security, privacy, and scalability. Blockchain consensus faces challenges too, for example, some popular \textit{proof--of--work}\endnote{Miller, A., \& LaViola, J. J. ``Anonymous Byzantine consensus from moderately--hard puzzles: A model for Bitcoin.'' \textit{Tech. Report}, CS-TR-14-01, UCF, (2014)}  algorithms result in energy waste\endnote{Baraniuk, C. ``Bitcoin energy use in Iceland set to overtake homes, says local firm''. \textit{BBC News} (accessed 24 October 2018) \url{https://www.bbc.com/news/technology-43030677}} whereas \textit{proof--of--stake}\endnote{King, S., Nadal, S. ``PPCoin: Peer-to-Peer Crypto-Currency with Proof-of-Stake'' (2012) (accessed September 2018) \url{https://archive.org/details/PPCoinPaper}} may result in resource or wealth concentration.\endnote{Zheng, Z., Xie, S., Dai, H., Chen X., Wang, H. ``An Overview of Blockchain Technology: Architecture, Consensus, and Future Trends.'' \textit{Proc. IEEE International Congress on Big Data}, 557-564, (2017)} As a result, fundamental redesign of blockchain architectures may be required to ensure the technology is capable of widespread acceptance whilst retaining its attractive qualities.\endnote{Croman K. \textit{et al.} ``On Scaling Decentralized Blockchains.'' \textit{Financial Cryptography and Data Security}, Lecture Notes in Computer Science, 9604, Springer, Berlin, Heidelberg (2016)}\endnote{Fedorov, A. K., Kiktenko, E. O., Lvovsky, A. I. ``Quantum computers put blockchain security at risk.'' \textit{Nature} \textbf{563} 465--467  (2018) doi:10.1038/d41586-018-07449-z} 

\begin{figure}[ht]
\centering
\includegraphics[width=\linewidth]{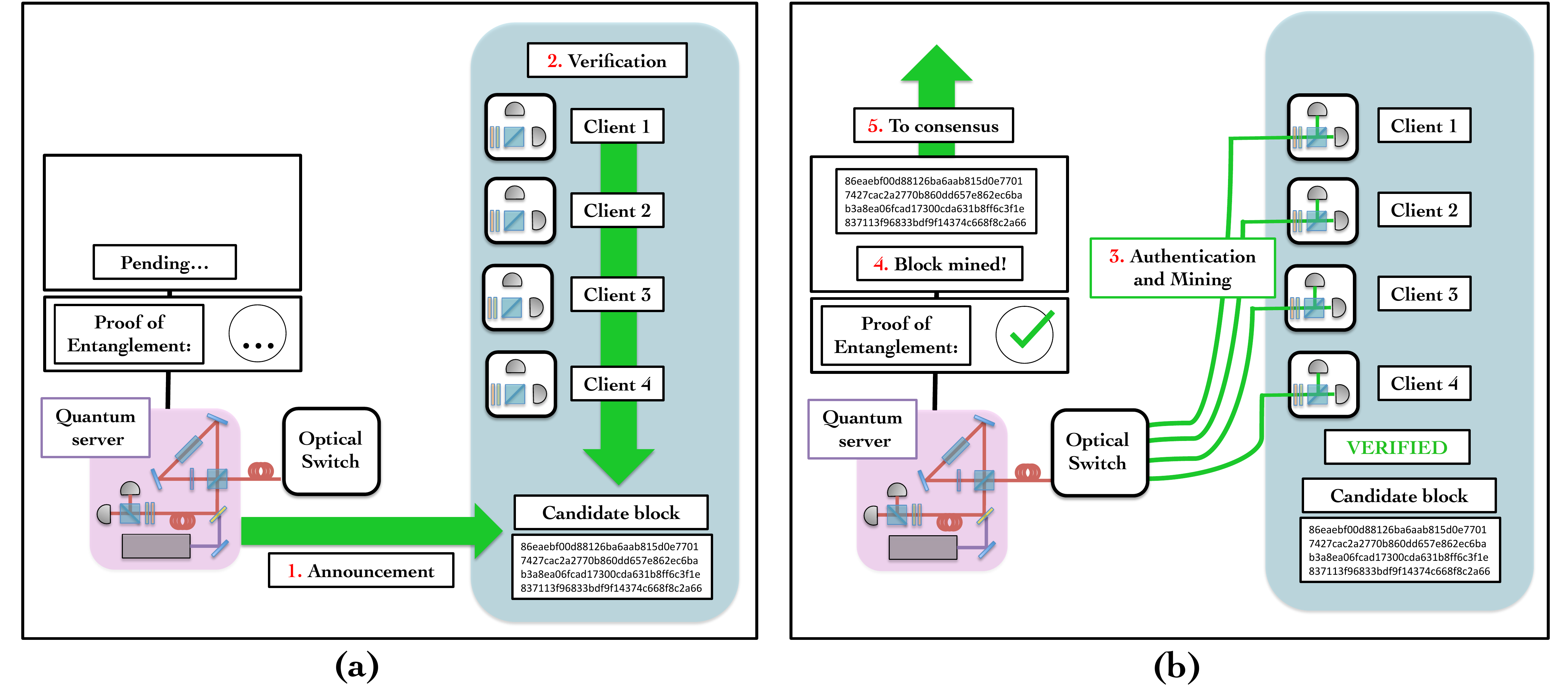}
\caption{\textbf{Overview of the mining protocol on a quantum--enabled consortium blockchain.} Clients and servers participate in an energy efficient interactive mining protocol for generating and committing entanglement as a resource towards securing a blockchain. In (a), servers announce candidate blocks to a pool of clients who verify blocks against verification criteria. In (b), authenticated clients may participate in block mining through an interactive protocol. Upon successful mining, the block is admitted into a consensus round for inclusion into the blockchain. Importantly, the physical network infrastructure is considered untrusted by the clients in the mining round, which includes the consortium of quantum servers. The authentication, mining, and consensus protocols are designed to detect malicious actions on behalf of the consortium and incentivise honest behaviours, such that quantum servers are held publicly accountable to act in the best interests of clients on the network. See main text for details.}
\label{fig:Q_POE}
\end{figure}

This work is a contribution in that direction, which explores mining new blocks on a quantum--enabled blockchain using light generated by energy efficient quantum optical devices (See Figure \ref{fig:Q_POE} for a conceptual overview). Quantum optical devices are a new type of network infrastructure for secure transmission and processing of quantum information. These devices offer advantages over traditional network infrastructures due to the fact that the information is transmitted, encoded, and decoded using physical systems which adhere to the laws of quantum physics. Quantum optical devices operate at the \textit{single photon} level, where a photon is the smallest permissible unit of light. Formally, a single bit of quantum information is called a quantum bit, or \textit{qubit}, and in the case of optical encodings, information is encoded onto a \textit{photonic qubit} using degrees of freedom which correspond to physical parameters of the photon (e.g. polarization, angular momentum, spatial mode, or a combination\endnote{Wang, X. L., \textit{et. al.} ``18--Qubit Entanglement with Six Photons' Three Degrees of Freedom.'' \textit{Phys. Rev. Lett} \textbf{120.26} 260502 (2018) doi: https://doi.org/10.1103/PhysRevLett.120.260502}). This optical encoding, when paired with the quantum properties of qubits, guarantees an inherent protection against eavesdropping,\endnote{Bennett, C, H., \& Brassard, G. ``Quantum Cryptography: Public key distribution and coin tossing.'' \textit{Proceedings of the IEEE International Conference on Computers, Systems, and Signal Processing}, 175, Bangalore, India (1984)}\endnote{Ekert, A. K. ``Quantum cryptography based on Bell's theorem.'' \textit{Phys. Rev. Lett.} \textbf{67}  661-663 (1991)}\endnote{Gisin, N., Ribordy, G., Tittel, W., Zbinden, H. ``Quantum Cryptography.'' \textit{Rev. Mod. Phys.} \textbf{74} 145  (2002)} affords compatibility with quantum computing,\endnote{Deutsch, D. ``Quantum theory, the Church--Turing principle and the universal quantum computer'' \textit{Proceedings of the Royal Society A}. \textbf{400} 1818 (1985)}\endnote{DiVincenzo, P. D. ``Quantum Computation.'' \textit{Science,} \textbf{270} 255--26 (1995)}\endnote{Monroe, C., Meekhof, D. M., King, B. E., Itano,W. M., \& Wineland, D. J. ``Demonstration of a Fundamental Quantum Logic Gate.'' \textit{Phys. Rev. Lett.} \textbf{75} 4714 (1995)}\endnote{Ekert, A., Jozsa, R. ``Quantum computation and Shor's factoring algorithm.'' \textit{Rev. Mod. Phys.} \textbf{68} 733 (1996)}\endnote{Ohlsson, N., Mohan, R. K., Kr\"oll, S. ``Quantum computer hardware based on rare--earth--ion--doped inorganic crystals.'' \textit{Optics Communications} \textbf{201.1} 71--77 (2002)} and permits secure information distribution and information teleportation on optical networks.\endnote{Pirandola, S., Eisert, J., Weedbrook, C., Furusawa, A., Braunstein, S. L. ``Advances in quantum teleportation.'' \textit{Nature Photonics}, \textbf{9} 641--652 (2015)}\endnote{N\"olleke, C., \textit{et. al.}``Efficient Teleportation Between Remote Single-Atom Quantum Memories.'' \textit{Phys. Rev. Lett.} \textbf{110} 140403 (2013)}\endnote{Valivarthi, R., \textit{et. al.} ``Quantum teleportation across a metropolitan fibre network.'' \textit{Nature Photonics} \textbf{10} 676--680 (2016)}\endnote{Xia, X. X.,  Sun, Q. C., Zhang, Q., \& Pan, J. W. ``Long distance quantum teleportation.'' \textit{Quantum Science and Technology} \textbf{3} 014012 (2017) \url{https://doi.org/10.1088/2058-9565/aa9baf}}\endnote{Bennett, C. H.,  Brassard, G., Cr\'epeau, C., Jozsa, R., Peres, A., Wootters, W. K. ``Teleporting an unknown quantum state via dual classical and Einstein--Podolsky--Rosen channels.'' \textit{Phys. Rev. Lett.} \textbf{70} 1895 (1993)}\endnote{Neumann, P., \textit{et. al.} ``Multipartite Entanglement Among Single Spins in Diamond.'' \textit{Science} \textbf{320.5881} 1326--1329 (2008)}  \\

Tests of this new infrastructure have proven the robustness of quantum optical devices in real--world scenarios. Some recent examples include quantum key distribution,\endnote{Salvail, L. \textit{et al}. ``Security of trusted repeater quantum key distribution networks.'' \textit{J. Comput. Sec.} \textbf{18} 61--87 (2010)}\endnote{Peev, M. \textit{et al}. ``The SECOQC quantum key distribution network in Vienna.'' \textit{New J. Phys.} \textbf{11} 075001 (2009)}\endnote{Stucki, D. \textit{et al}. ``Long--term performance of the Swiss quantum quantum key distribution network in a field environment.'' \textit{New J. Phys.} \textbf{13} 123001 (2011)}\endnote{Chen, T. Y. \textit{et al}. ``Field test of a practical secure communication network with decoy--state quantum cryptography.'' \textit{Opt. Express} \textbf{17} 6540--6549 (2009)}\endnote{Chen, T. Y. \textit{et al}. ``Metropolitan all--pass and inter--city quantum communication network.'' \textit{Opt. Express} \textbf{18} 27217--27225 (2010)}\\ \endnote{Kiktenko, E. O. \textit{et al}. ``Demonstration of a quantum key distribution network across urban fiber channels.'' \textit{arXiv:1705.07154} (2017)} quantum verification,\endnote{Shadbolt, P., Mathews, J. C. F., Laing, A., O'Brien, J. L. ``Testing foundations of quantum mechanics with photons.'' \textit{Nat. Phys.} \textbf{10} 278--286 (2014)}\endnote{Christensen, B. G. \textit{et. al}. ``Detection--loophole--free test of quantum nonlocality, and applications.'' \textit{Phys. Rev. Lett.} \textbf{111} 130406 (2013)}\endnote{Smith, D. H. \textit{et. al}. ``Conclusive quantum steering with superconducting transition--edge sensors.'' \textit{Nat. Comms.} \textbf{3} 625 (2012)}\endnote{Wittmann B. \textit{et. al.} ``Loophole--free Einstein--Podolsky--Rosen experiment via quantum steering.'' \textit{New J. Phys.} \textit{14} 053030 (2012)} and mobile quantum communication protocols,\endnote{Sasaki, M. ``Quantum networks: where should we be heading?'' \textit{Quantum Science and Technology} \textbf{2} 020501 (2017)}\endnote{Elmabrok, O., Razavi, M. ``Wireless quantum key distribution in indoor environments.'' \textit{Journal of the Optical Society of America B}, \textbf{35.2} 197--207 (2018)}\endnote{Chun, H., \textit{et al.} ``Handheld free space quantum key distribution with dynamic motion compensation.'' \textit{Optics Express} \textbf{25.6} 6784--6795 (2017)}\endnote{Zhang, P., \textit{et. al.} ``Reference--Frame--Independent Quantum Key Distribution Server with a Telecom Tether for an On--Chip Client.'' \textit{Phys. Rev. Lett.} \textbf{112} 130501 (2014)}\\ \endnote{Bourgoin, J. P., \textit{et. al.} ``Free--space quantum key distribution to a moving receiver.'' \textit{Optics Express} \textbf{23.26} 33437--33447 (2015)} all of which are compatible with existing telecommunication networks.\endnote{Toshiba Corporation. ``High--speed quantum cryptographic communications with key distribution speeds exceeding 10 Mbps in a real-world environment.'' \textit{Toshiba Press Release} (accessed 25th October 2018) \url{https://www.toshiba.co.jp/about/press/2018_09/pr1101.htm}}\endnote{Sasaki, M. \textit{et al}. ``Field test of quantum key distribution in the Tokyo QKD Network.'' \textit{Opt. Express} \textbf{19} 10387 (2011)}\endnote{Tysowski, P. K., Ling, X., L\"{u}tkenhaus, N., Mosca, M. ``The engineering of a scalable multi--site communications system utilizing quantum key distribution (QKD).'' \textit{Quantum Science and Technology} \textbf{3.2} (2018)} Here, we present a means of integrating quantum optical devices into a blockchain architecture, and deliver an interactive quantum protocol designed for energy efficient mining on a blockchain using light, namely using \textit{entanglement} as a resource for securing new blocks on a blockchain, which we term \textit{proof of entanglement} (PoE). Entanglement is a \textit{new kind of resource that is as real as energy,\endnote{Horodecki, R., Horodecki, P., Horodecki, M., Horodecki, K. ``Quantum Entanglement.'' \textit{Rev. Mod. Phys.} \textbf{81} 865 (2009)}} with applications in communication, cryptography, computation, and information distribution technologies. 

In this work, we elaborate upon results from two previous proof--in--principle experiments, and explore how the PoE mining protocol may be readily implemented with a Sagnac interferometer,\endnote{Fedrizzi, A., Herbst, T., Poppe, A., Jennewein, T., Zeilinger, A. ``A wavelength--tunable fiber--coupled source of narrowband entangled photons.'' \textit{Optics Express} \textbf{15.23} 15377--15386 (2007)}\endnote{Li, Y., Zhou, Z. Y., Ding, D. S., Shi, B. S. ``CW--pumped telecom band polarization entangled photon pair generation in a Sagnac interferometer.'' \textit{Optics Express} \textbf{23.22} 28792--28800 (2015)}\endnote{Jin, R. B. \textit{et. al.} ``Pulsed Sagnac polarization--entangled photon source with a PPKTP crystal at telecom wavelength.'' \textit{Optics Express} \textbf{22.10} 11498--11507 (2014)} functioning, in this case, as a \textit{quantum server}. The role of the quantum server is to generate entanglement, encoded in photonic qubits, which is sent to clients on the network via direct fibre link.  This approach to blockchain redesign with quantum \textit{compatibility} in mind contrasts starkly with contemporary trends in information technology, whereby developers resist emerging quantum technologies (like quantum computing) by implementing ``quantum--resistant" protocols and algorithms on traditional (i.e. non--quantum) platforms.\endnote{Ishai, Y., Kushilevitz, E., Ostrovsky, R., Sahai, A. ``Zero--Knowledge Proofs From Secure Multiparty Computation.'' \textit{SIAM J. Comput.} \textbf{39.3} 1121--1152 (2009) doi: 10.1137/080725398}\endnote{Torres, W. A. A., \textit{et al.} ``Post--Quantum One--Time Linkable Ring Signature and Application to Ring Confidential Transactions in Blockchain.'' In Susilo, W., Yang, G. (Ed.) \textit{Information Security and Privacy. ACISP 2018. Lecture Notes in Computer Science, vol. 10946.} Springer, Cham 558--576 (2018) \\ \url{https://doi.org/10.1007/978-3-319-93638-3_32}}\endnote{Gao X., Ding J., Liu J., Li, L. ``Post-Quantum Secure Remote Password Protocol from RLWE Problem.'' In Chen, X., Lin, D., Yung, M. (Ed.) \textit{Information Security and Cryptology. Inscrypt 2017. Lecture Notes in Computer Science, vol. 10726.} Springer, Cham 99--116 (2018) \\ \url{https://doi.org/10.1007/978-3-319-75160-3_8}} \blk By presenting an alternative blockchain architecture, we hope to promote further investigation into issues and advantages of novel quantum--enabled blockchain protocols,\endnote{Jogenfors, J. ``Breaking the Unbreakable: Exploiting Loopholes in Bell's Theorem to Hack Quantum Cryptography.''  \textit{PhD Thesis}, Linkoping University, Sweden (2017)}\endnote{Kalinin, K. P., Berloff, N. G. ``Blockchain platform with proof-of-work based on analog Hamiltonian optimisers.'' \textit{arXiv:1802.10091} (2018)}\endnote{Rajan, D., Visser, M. ``Quantum Blockchain using entanglement in time.'' \textit{arXiv:1804.05979} (2018)}\endnote{Ikeda, K. ``qBitcoin: A Peer--to--Peer Quantum Cash System.'' \textit{arXiv:1708.04955} (2017)}\endnote{Sapaev, D. \textit{et al}. ``Quantum--Assisted Blockchain.'' \textit{arXiv:1802.06763} (2018)}\endnote{Guan, J. Y. \textit{et. al.} ``Experimental preparation and verification of quantum money.'' \textit{arXiv:1709.05882} (2018)} and more generally, offer insight into the process of mapping traditional scientific computing technologies onto newly emerging quantum technologies.\endnote{Haylock, B., \textit{et. al.} ``Multiplexed Quantum Random Number Generation.'' \textit{arXiv:1801.06926} (2018)}\endnote{Lenzini, F., \textit{et. al.} ``Integrated photonic platform for quantum information with continuous variables.'' \textit{arXiv:1804.07435} (2018)}\endnote{Lenzini, F.,, \textit{et. al}. ``Active demultiplexing of single--photons from a solid--state source.'' \textit{Laser \& Photonics Reviews} \textbf{11} 1600297 (2017)}\endnote{Bogdanov, S. I., \textit{et. al.} ``Ultrabright Room--Temperature Sub--Nanosecond Emission from Single Nitrogen--Vacancy Centers Coupled to Nanopatch Antennas.'' \textit{Nano Lett.} \textbf{18.8} 4837--4844 (2018)}\endnote{Kandala, A., \textit{et. al.} ``Hardware--efficient variational quantum eigensolver for small molecules and quantum magnets'' \textit{Nature} \textbf{549} 242--246 (2017)}\endnote{Rancic, M., Hedges, M. P., Ahlefeldt, R. L., \& Sellars, M. J. ``Coherence time of over a second in a telecom--compatible quantum memory storage material''. \textit{Nature Physics} \textbf{14} 50--54 (2018)}

\section{Structure of paper}
In Section 3, we explore the connection between entanglement and trust. In Section 4, we detail the architecture and devices which support authentication, mining, and consensus on the quantum--enabled blockchain. In Section 5, we deliver our main result, the proof--of--entanglement mining protocol. In Section 6 we explore the results of two previous proof--in--principle experiments which support a preliminary energy efficiency analysis of the network and protocol. Section 7 offers a discussion and analysis of vulnerabilities. Section 8 concludes the work. 

\section{Entanglement and trust}
In 1935, researchers began exploring the notion that a pair of physical systems separated by vast distances might instantaneously influence one other. This idea sprouted from the mathematical formalism of the newly emerging quantum theory, and caused a great deal of controversy.\endnote{Wiseman, H. M. ``From Einstein's theorem to Bell's theorem: a history of quantum non-locality.'' \textit{Contemporary Physics} \textbf{47} 79--88 (2006)} At the time, the strangeness of the predicted phenomenon caused Einstein to reject the possibility of such occurrences, under the assertion that the emerging quantum theory must be incomplete.\endnote{Einstein, A., Podolsky, B., Rosen, N. ``Can quantum--mechanical description of physical reality be considered complete?'' \textit{Phys. Rev. Lett.} \textbf{47} 777 (1935)} Since then, experimental investigations into the phenomenon of entanglement have rigorously proven and characterised the effect,\endnote{Freedman, S. J., Clauser, J. F. ``Experimental Test of Local Hidden--Variable Theories.'' \textit{Phys. Rev. Lett.} \textbf{28} 938 (1972)}\endnote{Aspect, A., Dalibard, J., Roger, G. ``Experimental realization of Einstein--Podolsky--Rosen--Bohm \\Gedankenexperiment: a new violation of Bell's inequalities.'' \textit{Phys. Rev. Lett.} \textbf{49.2} 91--94 (1982) \\doi.org/10.1103/PhysRevLett.49.91}\endnote{Weihs, G., \textit{et. al.} ``Violation of Bell's Inequality under Strict Einstein Locality Conditions.'' \textit{Phys. Rev. Lett.} \textbf{81} 5039 (1998)}\endnote{Rowe, M. A., \textit{et. al.} ``Experimental Violation of a Bell's Inequality with Efficient Detection.'' \textit{Nature} \textbf{409} 791 (2001)}\endnote{Matsukevich, D. N., \textit{et. al.} ``Bell Inequality Violation with Two Remote Atomic Qubits.''  \textit{Phys. Rev. Lett.} \textbf{100} 150404 (2008)}\endnote{Ansmann, M., \textit{et. al.} ``Decoherence Dynamics of Complex Photon States Violation of Bell's inequality in Josephson phase qubits.'' \textit{Nature} \textbf{461} 504 (2009)}\endnote{Hofmann, J., \textit{et. al.} ``Heralded entanglement between widely separated atoms.'' \textit{Science} \textit{337} 72 (2012)}\endnote{Scheidl, T., \textit{et. al.} ``Violation of local realism with freedom of choice.'' \textit{Proc. Natl. Acad. Sci.} \textit{107} 19708 (2010)}\endnote{Aguero, M. B., Hnilo,A. A., Kovalsky, M. G. ``Time--resolved measurement of Bell inequalities and the coincidence loophole.'' \textit{Phys. Rev. A} \textbf{86} 052121 (2012)}\endnote{Giustina, M., \textit{et. al.} ``Bell violation using entangled photons without the fair--sampling assumption.'' \textit{Nature} \textbf{497} 227 (2013)}\endnote{Hensen, B., \textit{et. al.} ``Loophole--free Bell inequality violation using electron spins separated by 1.3 kilometres.'' \textit{Nature} \textbf{526} 682 (2015)}\endnote{Giustina, M., \textit{et. al}. ``Significant-Loophole--Free Test of Bell's Theorem with Entangled Photons.'' \textit{Phys. Rev. Lett.} \textbf{115} 250401 (2015)}\endnote{Shalm, L. K.,  \textit{et al}. ``Strong Loophole--Free Test of Local Realism.'' \textit{Phys. Rev. Lett.} \textbf{115} 250402 (2015)} and today, entanglement is placed as the ``missing link'' for the unification of quantum theory and gravity.\endnote{Susskind, L. ``Copenhagen vs Everett, Teleportation, and ER=EPR.'' \textit{Progress of Physics} \textbf{64.6} 551--564 (2016)}\endnote{Cao, C., Carroll, S. M., Michalakis, S. ``Space from Hilbert space: Recovering geometry from bulk entanglement.'' \textit{Phys. Rev. D} \textbf{95} 024031 (2017)}\endnote{Raamsdonk, M., V. ``Building up spacetime with quantum entanglement.'' \textit{General Relativity and Gravitation}, \textbf{42.10} 2323--2329 (2010)}\endnote{Susskind, L., Zhao, Y. ``Teleportation through the wormhole.'' \textit{Phys. Rev. D} \textbf{98} 046016 (2018)} Along the way, physical devices and algorithms have been developed in laboratories all over the world which have directly tested the existence and utility of the phenomenon. Recently, the devices involved in such tests have undergone rapid technological advancement,\endnote{Jiang, W. C., Lu, X., Zhang, J., Painter, O., Lin, Q. ``Silicon--chip source of bright photon pairs.'' \textit{Optics Express} \textbf{23.16} 20884--20904 (2015)}\endnote{Silverstone, J. W. \textit{et. al.} ``On--chip quantum interference between silicon photon--pair sources.'' \textit{Nature Photonics} \textbf{8} 104--108 (2014)}\endnote{Autebert, C., \textit{et. al.} ``Electrically Injected Source of Photon Pairs at Room Temperature: Device Performances and Entanglement Generation.'' \textit{European Conference on Lasers and Electro-Optics - EQEC} (2015)}\endnote{Horn, R., Abolghasem, P., Bijlani, B. J., Kang, D., Helmy, A. S., Weihs, G. ``Monolithic source of photon pairs.'' \textit{Phys Rev Lett.} \textbf{108.15} 153605 (2012)}\endnote{Davoyan, A., Atwater, H. ``Quantum nonlinear light emission in metamaterials: broadband Purcell enhancement of parametric downconversion.'' \textit{Optica} \textbf{5.5} 608--611 (2018)} and the majority of components, which typically consist of commercially available arrangements of lasers, mirrors, nonlinear crystals, optical detectors, and integrated photonic devices, have experienced reduced manufacturing costs and increased availability.\endnote{Joint Report: Interagency Working Group on Quantum Information Science of the Subcommittee on Physical Sciences. ``ADVANCING QUANTUM INFORMATION SCIENCE: NATIONAL CHALLENGES AND OPPORTUNITIES'' (accessed July 2018) \url{https://www.whitehouse.gov/sites/whitehouse.gov/files/images/Quantum_Info_Sci_Report_2016_07_22\%20final.pdf}} 

Interestingly, the interplay between the phenomenon of entanglement and the physical devices used in testing the existence of the phenomenon has also been the subject of intense research. These research outcomes have demonstrated that \textit{entanglement can be used as an indicator of device trustworthiness}, such that, under appropriate circumstances, if entanglement is observed, then a device's behaviour may be certified as trustworthy.\endnote{Mayers, D., Yao, A. ``Quantum cryptography with imperfect apparatus.'' \textit{IEEE Proceedings of the 39th Annual Symposium on Foundations of Computer Science} 503--509 (1998)}\endnote{Ac\'in, A., Brunner, N., Gisin, N., Massar, S., Pironio, S., Scarani, V. ``Device--independent security of quantum cryptography against collective attacks.'' \textit{Phys. Rev. Lett.} \textit{98} 230501 (2007)}\endnote{Pironio, S., Ac\'in, A., Brunner, N., Gisin, N., Massar, S., Scarani, V. ``Device--independent quantum key distribution secure against collective attacks.'' \textit{New Journal of Physics} \textbf{11} 045021 (2009)}\endnote{Branciard, C., Cavalcanti, E. G., Walborn, S. P., Scarani, V., Wiseman, H. M. ``One--sided device--independent quantum key distribution: Security, feasibility, and the connection with steering.'' \textit{Phys. Rev. A.} \textbf{85} 010301 (2012)}\endnote{Ac\'in, A., Gisin, N., Masanes, L. ``From Bell's Theorem to Secure Quantum Key Distribution.'' \textit{Phys. Rev. Lett.} \textbf{97} 120405 (2006)}\endnote{Liang, Y. C., Vertesi, T., Brunner, N. ``Semi--device--independent bounds on entanglement.'' \textit{Phys. Rev. A.} \textbf{83} 022108 (2011)} This is termed \textit{device--independent entanglement verification}, and is a powerful and practical application borne from research into the field of quantum information science. In these applications, initially untrusted pairs of non--communicating devices participate in a protocol which monitors the device's inputs and outputs, and, irrespective of the devices functionality, enacts a \textit{trust--test} on the device's behaviour based upon the experimental data which may guarantee a correct and therefore trusted functionality. The confirmation of trustworthiness certifies that the devices physically interact with an entangled resource in such a way that the resultant correlations observed in the experimental data cannot be recreated or emulated by \textit{any} arrangement of non--communicating analogue, digital, or quantum systems.\endnote{Brunner, N., Cavalcanti, D., Pironio, S., Scarani, V., Wehner, S. ``Bell nonlocality.'' \textit{Rev. Mod. Phys.} \textbf{86} 419 (2014)} Thus, the devices either yield a correct and secure result, or the trust--test detects that the devices are insecure (either working incorrectly or are under the control of a dishonest party), and any subsequent protocols (which may deal with valuable, sensitive, or private information) are aborted. Such protocols are highly desirable for practical implementation as they provide a higher level of security, unachievable by traditional systems. Here, we apply this technology towards developing blockchain--specific protocols which are energy efficient and secure against general adversaries. 

\section{Devices and architecture of the quantum--enabled blockchain}
\begin{figure}[ht]
\centering
\includegraphics[width=\linewidth]{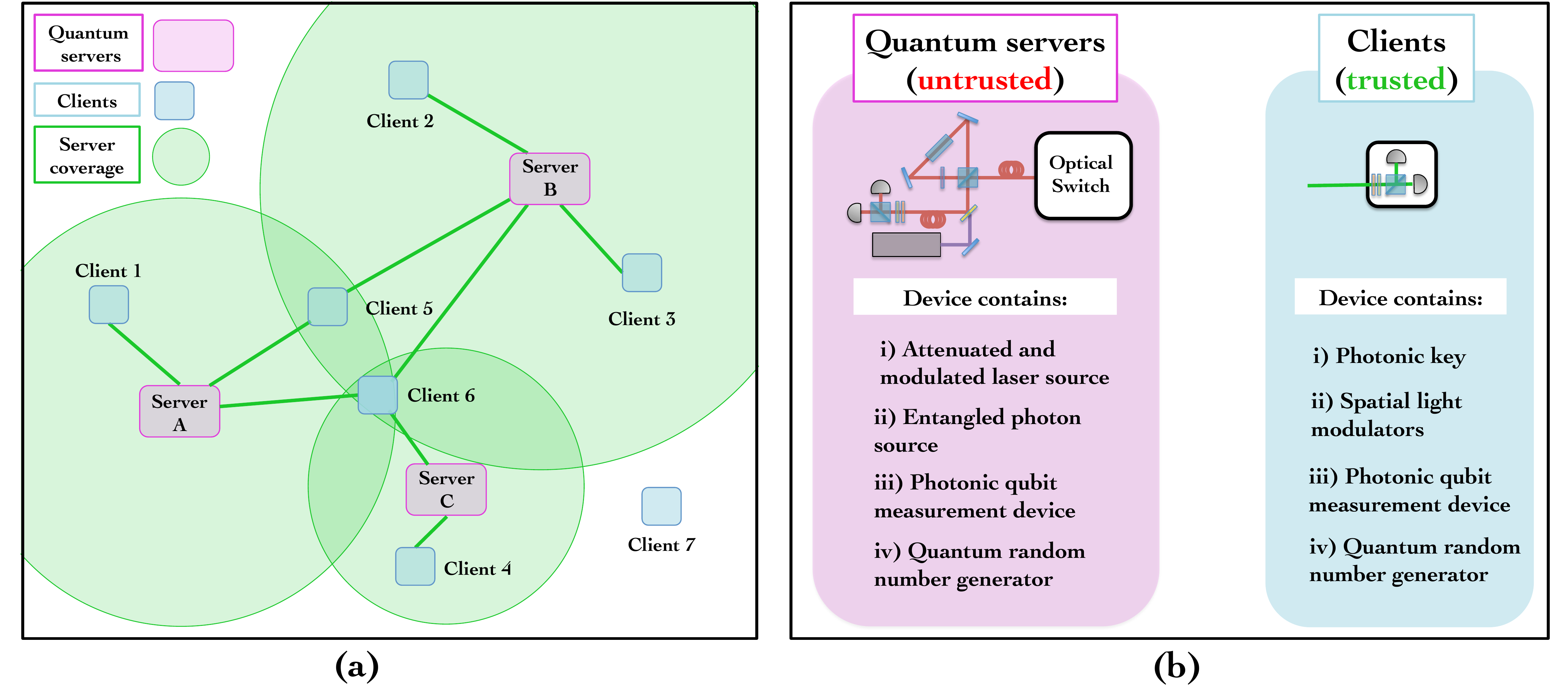}
\caption{\textbf{High--level overview of the quantum--enabled blockchain and devices.} a) Illustrates networking between an untrusted consortium of quantum servers (pink rectangles) and clients equipped with quantum modems (blue squares). Green links represent an insecure but authenticated classical communication channel and an authenticated quantum communication channel. Clients may access servers depending on the coverage map (green circles), with coverage radius depending upon infrastructure constraints. b) Presents an overview of the trust status and device subsystems required for interacting with the quantum--enabled blockchain. See main text for details. \blk}
\label{fig:Q_A}
\end{figure}

The architecture we propose consists of i) a network consortium of quantum servers, comprised of quantum optical devices which generate light for encoding quantum information, ii) a pool of clients whom operate \textit{quantum modems} for decoding quantum information stored in light, and iii) an optical fibre network which connects the client's quantum modems to the quantum servers. We assume a cryptographic scenario whereby the quantum servers and modems operate from secure environments which prevent unwanted information leakage. To communicate between one another, the servers and clients utilise insecure but authenticated classical channels and authenticated quantum channels (Figure \ref{fig:Q_A}(a)). The servers and clients are able to process classical information in a trusted way within their secure environment. Devices used for processing quantum information are assumed to operate in accordance with the laws of quantum physics, however in the mining round, no assumptions are made about the quantum servers or network infrastructure. An adversary may access (but not modify) classical information shared between the server and client, and may access and modify quantum information shared between them. An adversary is assumed complete knowledge of the authentication, mining, and consensus protocols, but does not have access to the random data generated by the client in their secure environment (except for information deduced from what they make public).

Figure \ref{fig:Q_A} (b) summarizes the network devices, device subsystems, and device's trust statuses, with quantum servers consisting of i) an attenuated and modulated laser source (for client authentication, described below), ii) a high--quality source of two--qubit photonic entanglement, iii) a device for performing measurements upon photonic qubits generated by the entanglement source, and iv) a source of randomness, for example, a quantum random number generator\endnote{Pironio, S., \textit{et. al.} ``Random numbers certified by Bell's theorem.'' \textit{Nature} \textbf{464} 1021--1024 (2010)} (QRNG). Quantum modems contain i) an embedded \textit{photonic key} and spatial light modulators (for client authentication, described below), ii) a measurement device for performing measurements upon photonic qubits, and iii) a QRNG, or a trusted and secure pseudo--random number generator.

Blockchain consensus protocols undertaken by a consortium of trusted entities have received little treatment in literature, but are of considerable interest in practice, with mainstream financial institutions actively exploring their use.\endnote{Shin, L. ``Bitcoin blockchain technology in financial services: How the disruption will play out.'' \textit{Forbes} (accessed November 2018) \textit{https://www.forbes.com/sites/laurashin/2015/09/14/bitcoin-\\blockchain-technology-in-financial-services-how-the-disruption-will-play-out/$\#$67a974bf5edf}} Traditionally, in a public blockchain everyone can participate in the consensus process. In a consortium blockchain, only a subset of ``nodes'', or in this case, quantum servers, are selected to participate in consensus. However, in the consortium blockchain architecture we propose, the consortium is considered \textit{untrusted} at the start of each round of block creation, and must be publicly elected by network clients through successful interactions in the mining round in order to qualify for the consensus round. In this way, clients trust only their own personal devices and the transparent network code. We now explore the details of block creation, which consists of an authentication round, mining round, and consensus round. 

\subsection{Authentication Round} 
The authentication round utilises a decoy--state variant of the \textit{quantum secure authentication} (QSA) protocol,\endnote{Goorden, S. A., Horstmann, M., Mosk, A. P., $\check{S}$kori{\'c}, B., Pinkse, P. W. H. ``Quantum-secure authentication of a physical unclonable key.'' \textit{Optica} \textbf{1} 421--424 (2014)}\endnote{Nikolopoulos, G.M., Diamanti, E. ``Continuous--variable quantum authentication of physical unclonable keys.'' \textit{Scientific Reports} \textbf{7} 46047 (2017)}\endnote{Skoric, B. ``Security analysis of Quantum--Readout PUFs in the case of challenge--estimation attacks.'' \textit{Cryptology ePrint Archive} (2013) (accessed 5 December) \url{https://eprint.iacr.org/2013/479}} allowing for hardware--based authentication of clients.The QSA protocol is a quantum cryptographic security primitive for the purposes of identification, authentication, read--proof key storage, key distribution, tamper detection, anti--counterfeiting, software--to--hardware binding, and trusted computing. Formally, the QSA protocol does not rely upon pre--shared secrets or seed keys, does not depend on the secrecy of stored data, does not depend on unproven mathematical assumptions, affords tamper resistance of devices, and is straightforward to implement with current technology (see Appendix A for additional information). The QSA protocol utilises photonic qubits (or more generally, quantum states of light), generated by a quantum server and sent, via a quantum communication channel, to a client--controlled quantum modem. The modems store an embedded photonic key, formally referred to as a \textit{physically unclonable key}, which are optically addressed and read using quantum states of light. Each photonic key is uniquely random by design, and  is impossible to clone, spoof, or copy.\endnote{Pappu, R., Recht, B., Taylor, J., Gershenfeld, N. ``Physical one-way functions.'' \textit{Science} \textbf{297} 2026 (2002)}\endnote{Pappu, R. ``Physical One-Way Functions.'' \textit{Ph.D. dissertation} Massachusetts Institute of Technology (2001)}\endnote{van Putten, E. G. ``Disorder--enhanced imaging with spatially controlled light.'' \textit{Ph.D. Thesis} University of Twente (2011)} Using quantum readout of physically unclonable keys guarantees accurate real--time verification and makes spoofing fundamentally difficult due to the no--cloning theorem of quantum physics\endnote{Wootters, W. K., Zurek, W. H. ``A single quantum cannot be cloned.'' \textit{Nature} \textbf{299} 802--803 (1982) doi: 10.1038/299802a0}. By utilising the quantum secure authentication protocol, vulnerability to the Sybil attack\endnote{Douceur, J. R. ``The sybil attack.'' \textit{Peer-to-peer Systems} 251--260 (2002)} is diminished as entry to the network is physically gated, incurring a resource cost in the case of multiple pseudonyms attributed to the same individual. As well as functioning as a secure hardware--based authenticator, the integrated photonic key may also serve as a physical address which can be uniquely attributed to a client on the network, allowing provision for dues paying or rewards to physical addresses that build a trustworthy reputation, and may be used in conjunction with a root seed for generating hierarchical deterministic wallets.\endnote{No author. ``Cryptocurrency Standards''. \textit{Tresor Wiki} (accessed October 2018) \url{https://wiki.trezor.io/Cryptocurrency$\_$standards}}

\subsection{Mining Round}
After authentication, the mining round utilises a variant of the \textit{EPR steering protocol},\endnote{Wiseman, H. M., Jones, S. J., Doherty, A. C. ``Steering, entanglement, nonlocality, and the EPR paradox.'' \textit{Phys. Rev. Lett.} \textbf{98} 140402 (2007)}\endnote{Evans, D. A., Cavalcanti, E. G., Wiseman, H. M. ``Loss--Tolerant Tests Of Einstein--Podolsky--Rosen Steering.'' \textit{Phys. Rev. A.} \textbf{88} 022106 (2013)}\endnote{Saunders, D. J., Jones, S. J., Wiseman, H. M., Pryde, G. J. ``Experimental EPR--steering using Bell--local states.'' \textit{Nat. Phys.} \textbf{76} 845--849 (2010)} termed the PoE protocol, which acts as an \textit{entanglement witness} and trust--test on the behaviour of a quantum server in the consortium (described in detail below). The PoE protocol is \textit{interactive}, and requires that the quantum server engage with clients on the network to generate the resource (entanglement or cumulative trust) for securing the blockchain, forming an example of \textit{resource--efficient mining}.\endnote{Zhang, F., Eyal, I., Escriva, R., Juels, A., van Renesse, R. ``REM: Resource--Efficient Mining for \\Blockchains.'' \textit{USENIX Security} (2017)} At the start of a new mining round, each quantum server announces a candidate block. Clients observe the announcements and, pending successful block verification, nominate themselves to participate in block mining. Clients are free to choose which servers to interact with (provided they reside within the region of server coverage) and work to elect servers into the voting round based upon i) the validity of the candidate block (agreement with block verification metrics), ii) the reputation of the quantum server (historical performance), iii) the throughput of the quantum server (i.e. rate of entanglement generation), and iv) connection availability (server proximity and reliability). Servers then assemble a team of unique clients, self--nominated for mining (termed \textit{miners}), to interactively generate and accumulate entanglement. Success in the mining round depends upon how much entanglement is generated by servers amongst their team of unique clients within the mining window. Quantum servers are thus incentivised to construct viable mining teams (in accordance with any construction metrics or regulations) and cultivate a reputation for honest behaviour. The subset of quantum servers in the consortium who ``win'' the mining round are then permitted to participate in the subsequent consensus round. 
 
\subsection{Consensus Round}
The results of the mining round determine which subset of quantum servers in the consortium are elected into the consensus round. The number of servers in the subset are determined by network protocol, only their identities change from round to round. We explore a consensus mechanism based upon a modified delegated  \textit{Byzantine fault tolerant protocol\endnote{Lamport, L., Shostak, R., Pease, M. ``The Byzantine generals problem.'' \textit{ACM Trans. Program. Lang. Syst.} \textbf{4} 382--401 (1982)}} (BFT) for securing new blocks and achieving decentralised consensus on the quantum--enabled blockchain. We choose a modified delegated BFT consensus protocol as it retains an element of decentralisation through interactive mining and circumvents the creation of branches off the main chain. Quantum servers that move into the consensus round are therefore always obligated to append newly mined blocks to the main chain with the most cumulative invested entanglement (or trust) recorded via PoE, and due to the mechanism of publicly electing delegates, the fate of the main chain is democratically guided as quantum servers must \textit{earn} the right to vote in the consensus round through successful interactions with clients. Consensus finality (also referred to as forward security\endnote{Decker, C., Seidel, J., Wattenhofer, R. ``Bitcoin meets strong consistency''. In \textit{17th International Conference on Distributed Computing and Networking} (2016)}) is satisfied by all BFT and state--machine replication protocols,\endnote{Schneier, F. B. ``Implementing fault--tolerant services using the state machine approach: A tutorial.'' \textit{ACM Comput. Surv.} \textbf{22.4} 299--319 (1990)} and BFT is resistant to server outages or arbitrarily long periods of asynchrony.\endnote{Dwork, C., Lynch, N., Stockmeyer, L. ``Consensus in the presence of partial synchrony.'' \textit{J. ACM} \textbf{35} (1988)} Here, consensus finality is a property which guarantees that blocks cannot be removed from the chain once added,\endnote{Vukoli{\'{c}}, M. ``The Quest for Scalable Blockchain Fabric: Proof-of-Work vs. BFT Replication.'' In Camenisch, J., Kesdo{\u{g}}an, D., \textit{Open Problems in Network Security}, Springer, Cham, 112--125, (2016).}\endnote{Xavier, D., Schiper, A., Urba{\'{n}}, P. ``Total order broadcast and multi- cast algorithms: Taxonomy and survey.'' \textit{ACM Comput. Surv.} \textbf{36.4} 372--421 (2004)} facilitating rapid confirmation and inclusions of transactions into the blockchain once candidate blocks have been successfully mined. \blk Upon initiation of the consensus round, one of the servers in the subset is nominated as a ``speaker'', with the remainder of servers presiding as ``delegates'', who are tasked with voting against the speaker's nomination. The speaker's nomination consists of the candidate block, mined by clients in the mining round, which the speaker presents on behalf of the clients for consideration by the consortium. Provided 66$\%$ of servers vote honestly, consensus will be reached. If the nomination and accompanying PoE are valid, then the block will be included in the chain. If the speaker's nomination is detected as being illegitimate (e.g. the speaker nominates a block with an invalid or falsified PoE), then the consensus round automatically starts anew, with a new speaker elected and the original speaker incurring a reputation loss and demotion to the role of delegate for that round. Section 7 covers potential attack vectors in more detail.

\section{The ``proof--of--entanglement'' mining protocol}
\subsection{Protocol background, assumptions, and characteristics} We now discuss the mining protocol in more detail. Candidate data blocks are announced to the network by quantum servers in the consortium, spreading to clients via the gossip protocol. The blocks must be compiled according to the network ruleset, consisting of a cryptographic hash of relevant information (e.g. recent transactions, code for smart contracts, digital media tags, etc.), which a quantum server recommends for inclusion into the blockchain data structure. Clients check the candidate blocks against key block verification criteria, and upon successful verification, will nominate themselves as miners for that block and wait to be placed in a mining team. 

As clients are drawn from the mining team, one by one, they will participate in the interactive mining protocol to test measurement data (generated between their quantum modem and the quantum server) for \textit{quantum correlations} which, importantly, cannot be faked (provided the clients satisfy criteria for \textit{one--sided} device--independence, described below). Tests of quantum correlations (or tests of ``quantumness'') arise due to physical interactions with entangled quantum systems, and are witnessed by mathematical inequalities. These inequalities are derived from constraints imposed upon probability distributions which quantify the degree of maximum classical correlation allowed under the assumptions of realism (from objective determinism) and locality (from relativity). These mathematical inequalities are well studied (for example, Clauser--Horne--Shimony--Holt inequalities\endnote{Clauser, J. F., Horne, M. A., Shimony, A., Holt, R. A. ``Proposed Experiment to Test Local Hidden--Variable Theories.'' \textit{Phys. Rev. Lett.} \textbf{23.15} 880--884 (1969)} or EPR--steering inequalities) and are known to act as an entanglement witness, allowing, in this case, a quantum server to prove honest distribution of entanglement to clients. For honest interactions, the entanglement generated in the mining round may then be ``invested'' towards securing the blockchain data structure in the subsequent consensus round. Dishonest interactions, which may attempt to fabricate entanglement by distributing unentangled qubits, falsify timestamps, or utilise Sybil or pre--mine attacks, are either detected or prohibited by the protocol, and are explored below in Section 7.

\begin{figure}[h!]
\begin{center}
\includegraphics[width=0.6\linewidth]{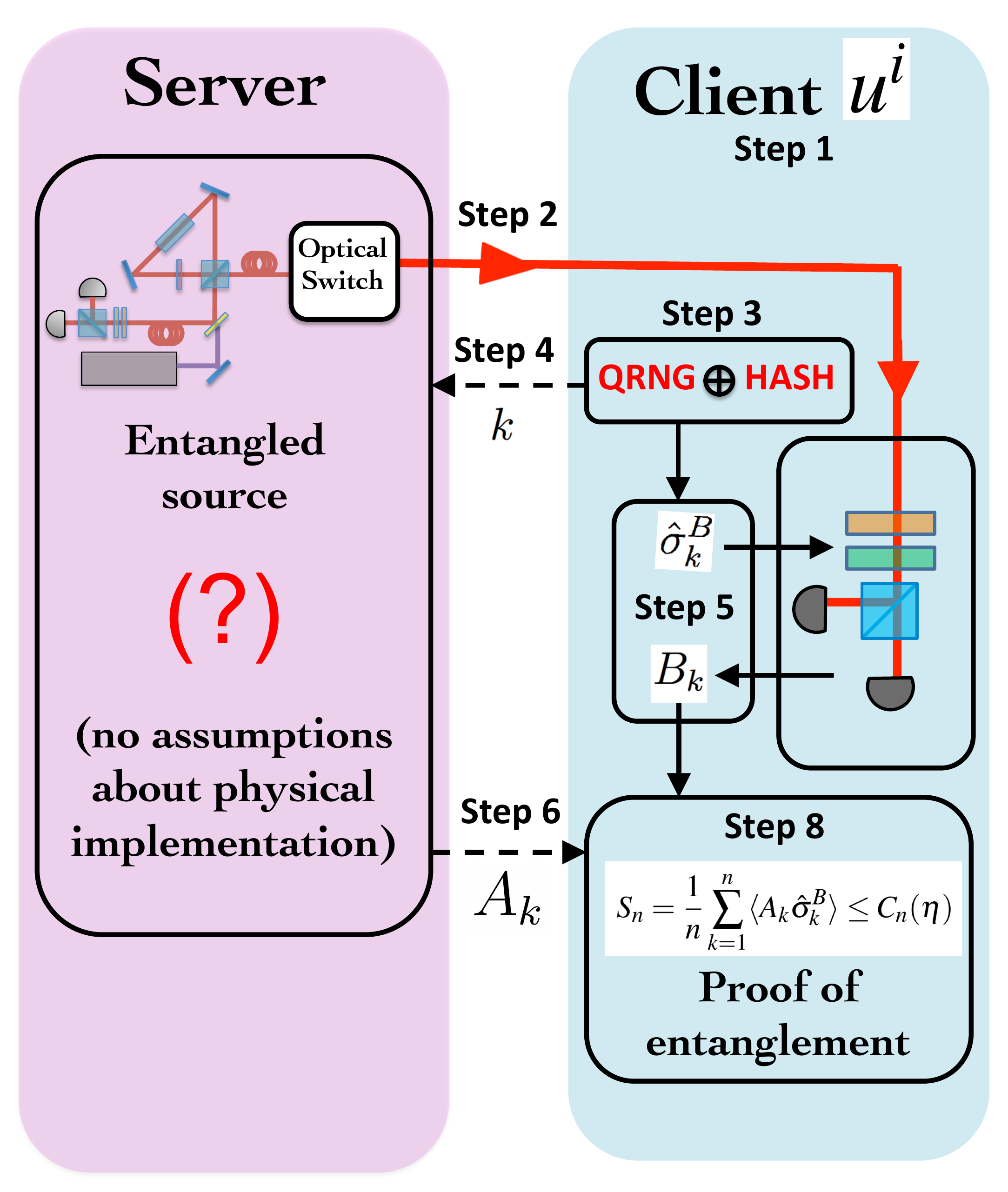}
\end{center}
\caption{\textbf{The PoE mining protocol, undertaken between a quantum server and client}. Purple shading (left--hand side) indicates the initially untrusted quantum server. Clear shading (right--hand side) indicates the client, who assumes the validity of quantum physics and trusts their measurement apparatus. Dashed arrows indicate classical communication channels between the quantum server and client. Solid arrows indicate quantum communication channels between the quantum server and client. Step numeration coincides with the steps of the PoE mining protocol outlined in the main text. In this example, the measurement device utilised by the client is assumed to perform polarisation measurements upon photonic qubits.}
\label{fig:BMP_EPR}
\end{figure}

Formally, in the case of the mining protocol delivered here, the trust--test performed against the quantum server will meet the requirements for one--sided device--independence provided that i) the data produced by the client's device is the result of measurements performed upon photonic qubits (this may be verified using measurement data from the authentication round), ii) the client's device does not leak information (a standard cryptographic assumption, which may be enforced with proper device design and tamper detection), iii) measurement inputs on the client side are uniquely generated at the time of measurement by random and unpredictable processes (enacted with a QRNG), and iv) detection efficiencies of relevant parties are properly considered and accounted for (in this case, via client--side tests of loss--tolerant EPR--steering inequalities). Appendix B elaborates on the specific safeguards required for observing device--independent entanglement witnesses. By ensuring the requirements for one--sided device independent entanglement verification are met, clients may interact with the blockchain whilst assuming nothing about the operation or makeup of external servers or infrastructure. Furthermore, clients trust the transparent network code, which formally acts as a third party, but in the case of EPR steering inequalities simply amounts to clients trusting themselves.\endnote{Jones, S. J., Wiseman, H. M., Doherty, A. C. ``Entanglement, Einstein--Podolsky--Rosen correlations, Bell nonlocality, and steering.'' \textit{Phys. Rev. A.} \textbf{76} 052116 (2007)} Put differently, clients with limited ``quantum powers'' (namely, the ability to perform measurements upon photonic qubits) can use the mining protocol to monitor and verify the actions of quantum servers such that any deviations from trustworthy behaviour are easily detectable.

\subsection{PoE mining protocol} One at a time, miners will be drawn from the mining team to participate in the interactive mining protocol, which proceeds as follows (Figure \ref{fig:BMP_EPR}):
\begin{quote}
\textbf{Step 1.} A miner $u^{i}$ is drawn from the mining team $\{u^{i}\}_{m}$, where $m$ is the total number of miners in the team. Each miner is accompanied by metadata which informs the quantum server of key parameters (e.g. physical address, reputation, measurement strategy). 

\textbf{Step 2.} The quantum server announces that it has transmitted a photonic qubit to the miner. 

\textbf{Step 3.} The miner generates a random measurement input for their device. This input is generated by reading the bitwise output from the client's QRNG. The output is combined under modulo addition with the bit value (or values, depending on measurement strategy) of the candidate block hash nominated by the quantum server for mining. This forms the measurement input, which tethers the hash of the candidate block to the PoE in a random way that can be validated at a later time by studying the measurement data. 

\textbf{Step 4.}  The client announces their choice of measurement input $k$ generated in Step 3 to the quantum server and performs a measurement upon the photonic qubit. 

\textbf{Step 5.} The client privately records the measurement outcome $B_{k}$. 

\textbf{Step 6.} The quantum server announces a measurement outcome $A_{k}$. Although no assumptions are made about the workings of the quantum server, correlated results will appear between the client's secretly recorded $B_{k}$ and the server's announced $A_{k}$, provided entanglement is present between the quantum server and client.

\textbf{Step 7.} Steps 2--6 are repeated until sufficient statistics are collected for the chosen measurement strategy. The measurement duration for each setting is tuned according to the client's detection efficiency, to ensure the minimisation of statistical errors. 

\textbf{Step 8.} The client calculates the \textit{heralding efficiency} $\eta$ of the quantum server and the EPR steering parameter $S_{n}$ (Equation 1, below) from the measurement data, and checks the calculated parameters against the inequality bound $C_{n}(\eta)$ required for entanglement verification (explored below). If the calculated EPR steering parameter does not exceed the bound required for entanglement verification, the network records a null event for that iteration (potentially incurring a reputation loss against the server) and the next miner is called in from Step 1. If a quantum server consistently fails to certify the presence of entanglement, it will be flagged for maintenance and may incur a reputation loss. 

\textbf{Step 9.} If the EPR steering parameter exceeds the bound required for entanglement, the inequality successfully witnesses entanglement. The resultant PoE and measurement data required for the verification  (e.g.  $S_{n}$, $\eta$) are appended to the block and attributed to the client $u^{i}$. 

\textbf{Step 10.} A new client is drawn from the team $\{u^{i}\}_{m}$, and Steps 1--9 are repeated until $m$ clients have witnessed entanglement generation of the quantum server. 
\end{quote}

The PoE uniquely ties the entanglement investment to the candidate block, and upon exhausting its team of miners (or upon closure of the mining window), the quantum server announces it's accumulation of PoE for that round, such that the results cannot be modified without re--doing the PoE. Thus, if the quantum server is elected as speaker in the voting round, the server is obligated to nominate it's newly mined candidate block. The PoE is easily verified simply by running the measurement data through the verification witness to ensure that the key parameters $S_{n}(\eta)$ and $\eta$ generated in the mining round observe entanglement. Here, the one--way functionality in resource expenditure, which is traditionally observed in proof--of--work systems, is also preserved via PoE, since generating the PoE requires physical interaction with an entangled quantum resource (``resource intensive''), but requires no physical interaction to certify the ``quantumness'' of the resultant verifier data (''resource light'').

\section{Energy cost and scalability analysis}
Demonstrating transmission of entanglement over a channel such as an optical fiber is important for real--world viability of the PoE protocol. Thus, we utilise the results of two previous proof--in--principle experiments implemented with a polarization Sagnac interferometer (Figure 4) to demonstrate the robustness and energy efficiency of the protocol. In the first experiment, a lossy channel was inserted into the entanglement source to test the validity of the one--sided device independent entanglement verification protocol,  even in cases of extreme photon loss.\endnote{Bennet, A. J., \textit{et. al.} ``Arbitrarily Loss--Tolerant Einstein--Podolsky--Rosen Steering Allowing a Demonstration over 1km of Optical Fiber with No Detection Loophole.'' \textit{Phys. Rev. X} \textbf{2} 031003 (2012)} This test characterised and addressed vulnerabilities in the entanglement verification protocol which might otherwise be used to exploit photon loss (on the ``server'' side) for fabricating entanglement. In the second experiment, a lossy channel was inserted between the output of the entanglement source (the ``server'') and the detection apparatus (the ``client''), closely mimicking a real--world client--server configuration.\endnote{Wollmann, S., Bennet, A. J., Walk, N., Wiseman, H. M., Pryde, G. J. ``Observation of Genuine One--Way Einstein--Podolsky--Rosen Steering.'' \textit{Phys. Rev. Lett.} \textbf{116} 160403 (2016)} In this second test, the client trusted their own measurement device (equivalent to a simplified quantum modem), and holds the entanglement source (equivalent to a simplified quantum server) accountable to demonstrate honest behaviour by overcoming the inequality constraints placed upon the server's reported efficiencies tested in the first experiment. \\

\begin{figure}[h!]
\begin{center}
\includegraphics[width=\linewidth]{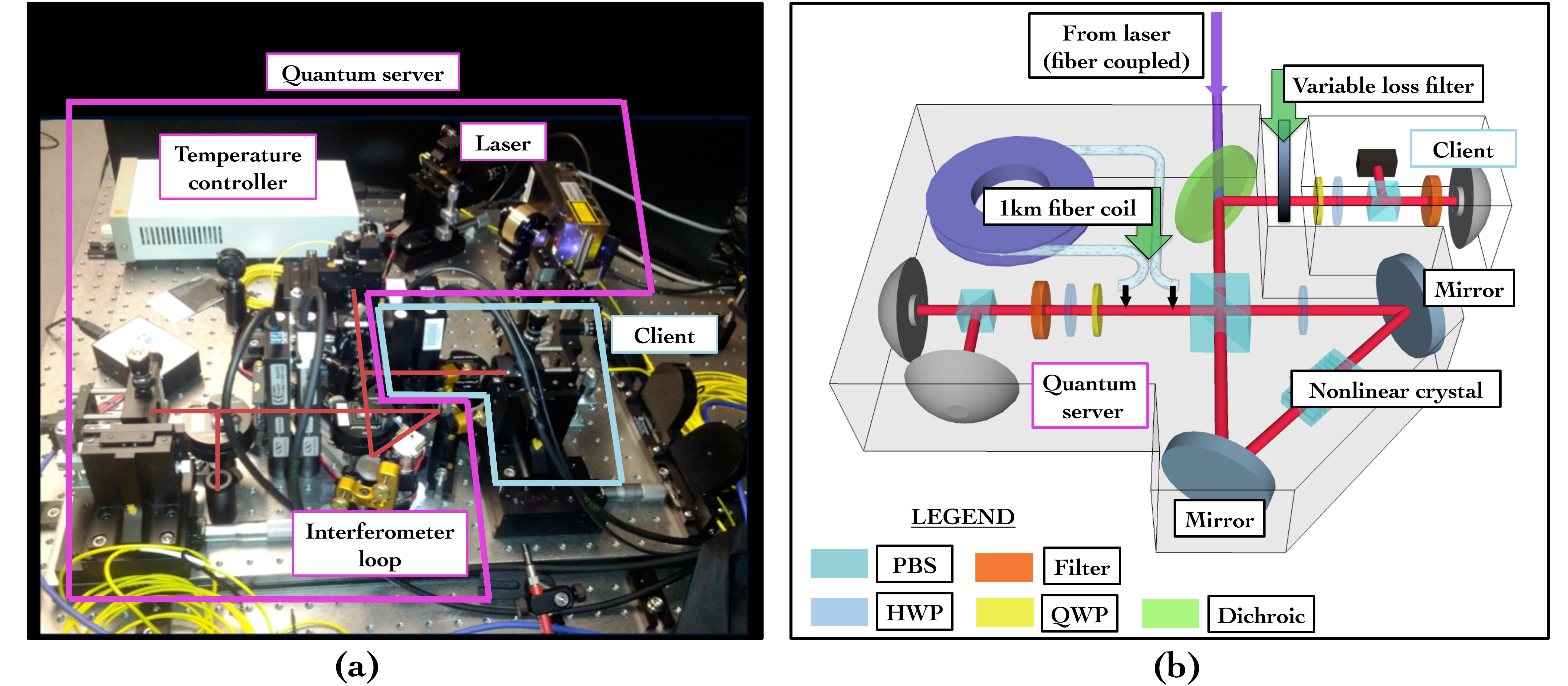}
\end{center}
\caption{\textbf{An entanglement source, based upon a polarization Sagnac interferometer, equivalent to a quantum server.} (a) A photograph of the experimental apparatus used in previous experimental demonstrations (testing the loss--tolerance and viability of the entanglement verification protocols) which may be implemented as a quantum server and quantum modem to implement the PoE mining protocol. The device is compact, fitting on a 600x600mm board. (b) The server and client devices are energy efficient, consisting mostly of passive optical elements, which include polarising beam splitters (PBS), wavelength filters, silvered and dichroic mirrors, an optical fiber line, a variable loss neutral--density filter, and half waveplates (HWP) and quarter waveplates (QWP) for implementing the polarisation measurements required to test entanglement witnesses based upon loss--tolerant EPR steering inequalities (see main text for details).}
\label{fig:Sagnac1}
\end{figure}

In implementing the PoE protocol, the client trusts their own device, so that protocol iterations which fail to detect a photonic qubit are safely discarded. However, the client cannot trust any claims the server makes about the propagation losses or the efficiency of its detectors. In particular, the client does not trust the server's claims about how often it sees a photonic qubit, conditional on the client's detecting one. Rather, the client makes use only of the server's heralding efficiency $\eta$: the probability that the server \textit{heralds} the client's result by declaring a measurement outcome (or equivalently, nonzero prediction) $A_{k}$ for it. The quantity $\eta$ is determined by the client wholly from the publicly announced frequencies of events to which the client (and the network) has direct access. The key result is that the client can circumvent loss--dependent vulnerabilities, even in the presence of arbitrarily high loss, by computing the witness $S_{n}$ based upon the shared correlations from the inequality

\begin{equation}
S_{n} = \frac{1}{n}\sum^{n}_{k=1}\langle A_{k} \hat{\sigma}^{B}_{k}\rangle\leq C_{n}(\eta)
\end{equation}

and observing that the inequality is \textit{violated}, thereby signifying entanglement. Here, the bound $C_{n}(\eta)$ is derived by the client, who in doing so, utilises the optimal cheating strategy available to the server for the observed efficiency and pre--determined measurement strategy $\{\hat{\sigma}^{B}_{k}\}_{n}$. The client's quantum mechanical measurement operator, denoted  $\hat{\sigma}^{B}_{k}$, defines their randomly selected choice of measurement setting $k$ drawn from the pre--determined set. In our experimental demonstrations, these operators define projective polarization measurements, which in the formalism of optical quantum theory, map an incident qubit (optically encoded in polarization) onto a polarization measurement basis defined by the transmission and reflection ports of a polarizing beam splitter. 

We utilise the results of the two previous experiments as an analysis and simulation of mining via PoE. To ensure a good signal--to--noise ratio of detection events on the client side, the photonic qubit detection rate was optimised to between $\approx$3000--6000 detection events per second. The total energy consumption used by the devices in the entanglement verification stage is estimated at 86W for the quantum server, which consisted of (not including passive optical elements, which do not consume electricity) two photon detectors (Perkin Elmer SPCM-AQR-14-FC), one temperature control unit (Thorlabs TEC200C), two waveplate controllers (Newport PR50CC), one continuous--wave laser (Toptica iBeam 405nm), and one coincidence electronics unit (UQDevices FPGA). The energy consumption for the client is estimated at 19W, consisting of one detector (Perkin Elmer SPCM-AQR-14-FC) and two waveplate controllers (Newport PR50CC). This measure of energy consumption excludes the energy costs associated with the PCs required to run the equipment, since PCs are assumed to be a pre-existing component in a server/client infrastructure. The devices considered in our analysis are simply those needed to upgrade an existing digital system infrastructure to include quantum capability. At any given moment, the rate of energy delivery per quantum server running the interactive PoE protocol with a client is estimated at 105W. 

For our energy cost analysis, assuming continuous operation over a 24 hour period and a 10 minute block mining window, the PoE protocol is predicted to service between 288 to 2880 clients (2 to 20 clients per mining team, randomly selected) who interactively mine 144 new blocks. Assuming energy efficient quantum servers with entanglement generation rates between those we tested ($3000$ entangled photon pairs per second) up to quantum servers based upon the latest integrated photonic technologies ($6\times10^{7}$ photon pairs per second\endnote{Jiang, W. C., Lu,X., Zhang, J., Painter, O., Lin, Q. ``Silicon--chip source of bright photon pairs.'' \textit{Optics Express} \textbf{23.16} 20884 (2015) doi: 10.1364/OE.23.020884}), each server can be expected to reliably service clients via direct fibre link in a 1km to 50km radius (assuming optical fiber transmission losses of 0.5dB/km, coupling losses of 80$\%$, and detection losses of $50\%$). Assuming a consortium testnet of 2000 quantum servers, the energy consumption based on interactive mining via the PoE protocol is of order 210kW. Using specially designed hardware, renewable energy sources, and further optimising server throughput will help to keep energy consumption low, with the added bonus of forward--compatibility with a quantum internet and optical quantum computing technologies. 

\section{Discussion and analysis of vulnerabilities}
We note that in a real--world implementation, the client must choose their measurement settings independently from one iteration of the mining protocol to the next, and safeguards must be in place to ensure that no information escapes from the client's devices unless they allow it. As these are standard assumptions in cryptographic proofs, these safeguards were not imposed in the previous experimental demonstrations. In addition, since we (the experimenters) controlled the server's implementation of honest or dishonest strategies, there was no need to enforce a strict time ordering of events. In a field deployment, the protocol requires that the client only accepts the server's announcements $A_{k}$ after the client measures their photonic qubit and declares their measurement setting $k$, in accordance with the PoE protocol described. 

\begin{figure}[h]
\centering
\includegraphics[width=0.55\linewidth]{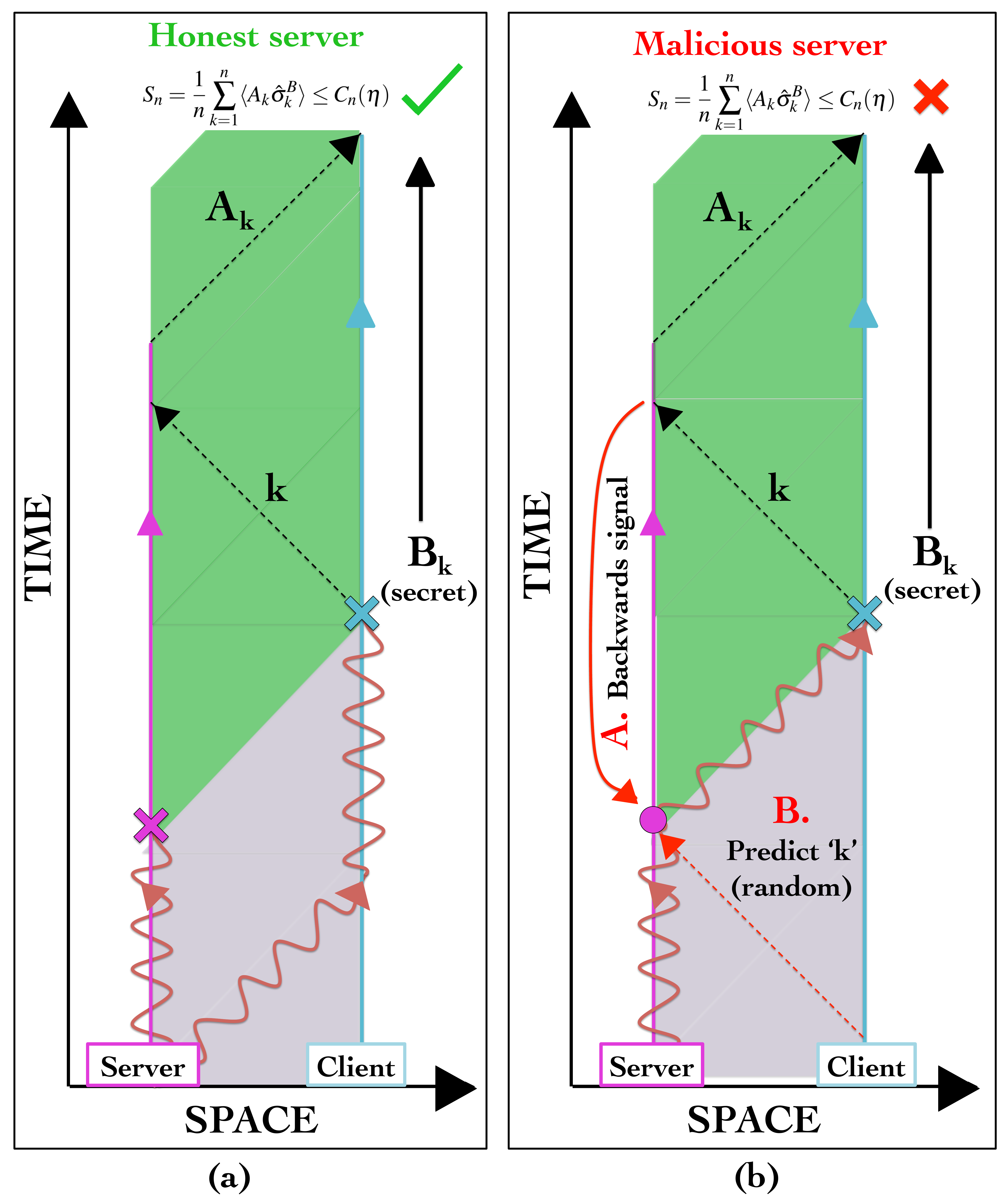}
\caption{\textbf{Spacetime diagrams for mining via PoE in the case of an honest and dishonest quantum server.} Red squiggles represent photonic qubits, which in (a), are entangled. Measurements are denoted using pink (quantum server) and blue (client) crosses. Classical data is communicated with dashed lines in accordance with the mining protocol (main text). Secret data is communicated with solid lines. Purple shading represents times during which the quantum server is able to transmit a photonic qubit, within the limitations of the protocol, to the client. Green shading represents times during which the server may announce measurement outcomes $A_{k}$ (recorded by the client after they perform their measurement). In (b), so--called cheating strategies are considered which permit false announcements, falsified timestamps, and fabricated qubit encodings (pink disk), which a server might utilise in an attempt to fabricate entanglement. Successful attempts require violating backwards non--signalling (A) or measurement--independence (B). See main text for details. \blk}
\label{fig:Q_ST}
\end{figure}

This latter requirement circumvents a potential vulnerability, exploitable by a quantum server, based upon falsified announcements of entanglement generation and fabricated qubit encodings which might allow a server to fabricate entanglement (termed a \textit{cheating strategy} in literature). However, successfully exploiting these vulnerabilities requires that the server either: a) knows the variables determining the behaviour of the client's QRNG, or b) violates \textit{backward non--signalling}\endnote{Friedman, R. A., H\"{a}nggi, E., Ta--Shma, A. ``Towards the Impossibility of Non--Signalling Privacy Amplification from Time--Like Ordering Constraints.'' \textit{arXiv} (2012) (accessed 14 January 2019) \url{https://arxiv.org/abs/1205.3736v1}}, whereby the server is able to send encoding information backwards in time to a photonic qubit (Figure \ref{fig:Q_ST}). In the case of the former (a), access to these variables was proven impossible in 1964 by Bell,\endnote{Bell, J. S. ``On the Einstein--Podolsky--Rosen paradox.'' \textit{Physics (Long Island City, N.Y.)} \textbf{1} 195 (1964)} who demonstrated that such variables do not exist for quantum systems, meaning no adversary can predict measurement outcomes prior to a measurement being made (called measurement--independence, see Appendix B). And in the latter case  (b), signalling encoding information backwards in time for use in a cheating strategy based upon fabricated qubit encodings amounts to a violation of the standard assumption of locality (in relativistic mechanics, faster than light signalling and signalling backwards in time are indistinguishable). Thus, provided no information leaks out of the client's device, and the client's device performs measurements upon photonic qubits to test loss--tolerant EPR steering inequalities, the PoE will generate data that cannot be simulated or spoofed by a quantum server. This distinction of ``quantumness'' means that, as long as an inequality violation is observed, we have the guarauntee, independently of any implementation details on the server side, that the two systems measured by the client and server are entangled. 

The risk of a traditional pre--mine attack against the network is diminished by infrastructure cost due to the interactive nature of the mining protocol. Because real data generated in a genuine round of block creation can be distinguished from attempted simulations of the required data, an attacker planning a pre--mine attack will need to collude with a large pool of clients to stage a Sybil attack, with each client needing their own quantum modem and photonic key. Furthermore, if mining pool construction demands a random selection of clients, an attacker would need to wait for an optimal mining team assemblage. This means an attacker planning a Sybil attack would be better incentivised to act as an honest participant. By ensuring the identities of quantum server operators/administrators are publicly known and certified prior to entry to the consortium and for participation in consensus via BFT, double spend attacks may be detected, recorded, and prevented in the consensus round (also at the start of the round of block creation through block verification by clients), provided two thirds of quantum servers on the network are acting honestly and serving the best interests of the network. Heavy increases in the throughput of a quantum server (in terms of entanglement generation rates) allows servers to overcome transmission losses more easily whilst remaining publicly accountable and limited in their capacity to arbitrarily command network consensus. This allows for an increased area of coverage and/or greater flexibility in constructing mining teams, which due to the interactive nature of the mining protocol, benefits clients too. 

Other factors which affect mining rates include detection efficiencies, transmission loss, coupling efficiencies, measurement strategy, and the switching rate between measurement settings. To optimise network performance and ensure fair competition between servers, a provider might impose reward restrictions or sanctions against a subset of high--performance quantum servers, or choose to implement composition rules for constructing mining teams, opting for completely general mining (maximum competition and freedom amongst servers), standardised mining (some competition and restricted freedoms), or restricted mining (collaborative and regulated). Servers which perform well may build a strong reputation and potentially be rewarded, depending on the network objectives and ruleset. \\ 

As a final commentary, at a high level, the quantum--enabled blockchain architecture we consider aims to incorporate four systems which work together to enable and incentivize trustworthy behaviour, these being\endnote{Schneier, B. \textit{Liars and Outliers. Enabling the Trust that Society Needs to Thrive} Indianapolis, Indiana: John Wiley \& Sons, Inc. 61--123 (2012)}: morals, reputation, institutions, and security. The first two systems, morals and reputation, are supported through a system of peer--to--peer trust, whereby clients and servers come to trust one another through public interactions and build a reputation for honest behaviour.\endnote{Werbach, K. \textit{The Blockchain and the New Architecture of Trust} Cambridge, MA: MIT Press 25--28 (2018)} The third system, institutions, traditionally have rules, contracts, and laws which formalize reputation and encourage people to behave according to the group norm (and enact sanctions against those who do not). The institutional trust system, supplanted by transparent network code, allows parties that don't trust each other (i.e. server/client pairs) to enter into an agreement through mutual trust in an overarching system of rules, laws, or logic that will help resolve disputes (implementing, for instance, protocols based upon the validity of relativity and quantum physics). The fourth system for incentivizing trustworthy behaviour, security systems, encompasses the wide variety of available security technologies used for detecting tampering, detecting device faults, detecting eavesdropping, ensuring records are immutable and auditable, ensuring anti-counterfeiting, and ensuring privacy through secure authentication and encryption technologies.

Blockchain technology allows developers to shift some of the trust in people and institutions to trust in technology. However, any evaluation of the security of the system has to take the whole socio-technical system into account\endnote{Schneier, B. ``Blockchain and Trust.'' \textit{Schneier On Security} (accessed 16 February 2019) \url{https://www.schneier.com/blog/archives/2019/02/blockchain_and_.html}}. In architectures which carry financial and social implications or consider social dynamics, it is not enough to focus solely on the technology and ignore the social element, as network trust can't be entirely replaced by algorithms and protocols. There is a social system too, where decisions in that system may be informed by accountability and reputation. Therefore, in this work, we aim to establish a network architecture whereby use of the network strengthens existing trust relationships. We aim to implement a trust model whereby abuses of trust do not incur irreversible damages to those exploited, but do incur reputation penalties for offenders. We argue that the system architecture we explore, were it not to utilise a blockchain architecture, would lack the flexibility required to support a broader social context which may utilise metrics for decision making and for quantifying accountability and reputation.\blk

\section{Conclusion}
Here we have presented an interactive mining protocol for energy efficient mining on a quantum--enabled consortium blockchain. The architecture is designed to ensure that the consortium behaves in the best interest of the network, removing the need for trust in server operators or network administrators in the mining round. Instead, clients on the network need only trust the network code and their own devices, and the consortium is held accountable to act in accordance with the network ruleset. We explore the use of a quantum--secure authentication protocol for authenticating clients, hardware, and communication channels, for physical addressing, for prevention against a Sybil attack and for tamper detection of devices. The tamper detection supports the assumption that devices do not leak information and that devices function according to the rules of quantum mechanics, which permits the use of entanglement as a resource for securing the blockchain data structure through an interactive mining protocol. 

Many potential improvements remain. The possibilities and ultimate limitations of a quantum--enabled blockchain remain largely unexplored, including benchmarks for transaction throughput, exploring alternative consensus mechanisms (for instance, those which allow for chain forks), alternative mining protocols, latency and storage optimisations, optimal signing and authentication schemes, optimal mining team construction algorithms, privacy, and a deep analysis of network vulnerabilities. We note that recent demonstrations of quantum--secured blockchain networks demonstrate the application of quantum digital signatures\endnote{Kiktenko, E. O., \textit{et. al.} ``Quantum-secured blockchain.'' \textit{Quantum Sci. Technol.} \textbf{3} 035004 (2018) doi:10.1088/2058-9565/aabc6b}, however, the architecture requires an additional infrastructure requirement, namely, that all quantum servers share quantum communication channels (i.e. optic fiber links) between one another, and furthermore does not address consensus. \blk In future works, we plan to expand upon our proof--in--principle experimental demonstrations to include promotion into a voting round and quantum secure authentication capabilities, and additionally emphasise the scalability and compatibility of the quantum--enabled blockchain with current telecommunication infrastructure and emerging quantum technologies. 

Looking forwards, there is a promising outlook for quantum networks and protocols,\endnote{Gottesman, D., Chuang, I. ``Quantum Digital Signatures.'' \textit{arXiv:quant-ph/0105032} (2001)}\\\endnote{Lo, H. K., Curty, M., Qi, B. ``Measurement--Device--Independent Quantum Key Distribution.'' \textit{Phys. Rev. Lett.}, \textbf{108} 130503 (2012)}\endnote{Smania, M., Elhassan, A. M., Tavakoli, A., Bourennane, M. ``Experimental quantum multiparty communication protocols.'' \textit{NPJ Quantum Information} \textbf{2} 16010 (2016)}\endnote{Fitzi, M., Gisin, N., Maurer, U. ``Quantum solution to the Byzantine agreement problem.'' \textit{Phys. Rev. Lett.} \textbf{87} 217901 (2001)}\endnote{Bancal, J D., Brunner, N., Gisin, N.,  Liang, Y. C. ``Detecting Genuine Multipartite Quantum Nonlocality: A Simple Approach and Generalization to Arbitrary Dimensions.'' \textit{Phys. Rev. Lett.} \textbf{106} 020405 (2011)}\endnote{Brunner, N., Sharam, J., V\'ertesi, T. ``Testing the Structure of Multipartite Entanglement with Bell Inequalities.'' \textit{Phys. Rev. Lett.} \textbf{108} 110501 (2012)}\endnote{Kocsis, S., Hall, M. J. W., Bennet, A. J., Saunders, D. J., Pryde, G. J. ''Experimental measurement--device--independent verification of quantum steering.'' \textit{Nature Comms.} \textbf{6} 5886 (2015) \\doi: https://doi.org/10.1038/ncomms6886} especially those which use limited infrastructures and supplement quantum computing. For instance, recent work on verifiable delegated quantum protocols\endnote{Reichardt, B. W., Unger, F., Vazirani, U. ``A classical leash for a quantum system: Command of quantum systems via rigidity of CHSH games'' \textit{arXiv:1209.0448} (2012)}\endnote{Gheorghiu, A., Wallden, P., Kashefi, E. ``Rigidity of quantum steering and one--sided device--independent verifiable quantum computation.'' \textit{New Journal of Physics}, \textbf{19} (2017)}\endnote{Mahadev, U. ``Classical Homomorphic Encryption for Quantum Circuits.'' \textit{arXiv} (accessed 15 December 2018) \url{https://arxiv.org/abs/1708.02130v4}} explore how clients with limited ``quantum powers'' -- namely, the ability to perform measurements upon qubits, might interact with quantum technologies to verify the trustworthiness of data outputs. An example in the case of quantum computation is a client enacting a verifiable delegation protocol to ensure that an untrusted cloud--based quantum computer correctly performs encrypted quantum computations on behalf of a client. It may be useful to explore more deeply the utility of these kinds of protocols against untrusted quantum servers, and more generally to explore the utility of quantum modems for enacting verifiable delegated quantum protocols against quantum--enabled network infrastructures. 

The quantum--enabled blockchain architecture we describe requires a fibre optic infrastructure, making the network challenging to implement on a large scale. However, the ability to witness entanglement in the presence of large losses opens new possibilities for security in long-range transmission of photonic entanglement over optical fiber, through free space\endnote{Ursin, R. \textit{et. al.} ``Entanglement--Based Quantum Communication over 144 km.'' \textit{Nature Phys.} \textbf{3} 481--486 (2007) doi: 10.1038/nphys629}, or to a satellite.\endnote{Aspelmeyer, M., Jennewein, T., Pfennigbauer, M., Leeb, W. R., Zeilinger, A. ``Long--Distance Quantum Communication with Entangled Photons Using Satellites.'' \textit{IEEE J. Sel. Top. Quantum Electron.} \textbf{9.6} 1541--1551 (2003) doi: 10.1109/JSTQE.2003.820918} The architecture we explore opens further avenues for quantum technologies in the blockchain space, with initial applications perhaps best suited for use in enterprise, due to the consortium architecture and ease of retrofitting short--to--medium range fibre optic interconnects. We expect this infrastructure would be particularly interesting to those with a vision for compatibility with quantum computing, the quantum internet, and the use of quantum cryptographic encryption protocols on a blockchain.

\theendnotes

\newpage

\appendix
 \setcounter{section}{0}
 \renewcommand\appendixname{Supplement}
\section{Quantum Secure Authentication}
\subsection{Background} In a real--world setting, the network ruleset for client authentication should meet modern cybersecurity standards for globally connected devices,\endnote{No author. ``Framework for Improving Critical Infrastructure Cybersecurity.'' \textit{National Institute of Standards and Technology} (accessed October 2018) \\\url{https://nvlpubs.nist.gov/nistpubs/CSWP/NIST.CSWP.04162018.pdf}} which naturally includes quantum optical devices embedded into a quantum--enabled blockchain. As a general rule, authentication mode should ensure\endnote{``Security Tenets for Life Critical Embedded Systems.'' \textit{Department of Homeland Security} (accessed October 2018) \textit{https://www.dhs.gov/sites/default/files/publications/security-tenets-lces-paper-11-20-15-508.pdf}}: 

\begin{quote}
i) All interactions between devices are mutually authenticated; \\
ii) Continuous authentication is used when feasible and appropriate; \\
iii) All communications between devices is encrypted; \\
iv) Devices never trust unauthenticated data or code during boot--time; \\
v) Devices are never permitted to run unauthorised code; \\
vi) Devices never trust unauthenticated data during run--time; and\\
vii) When used, cryptographic keys are protected, or are one--time use only (depending on protocol). 
\end{quote}

The integrity of QSA is guaranteed provided that\endnote{Goorden, S. A., Horstmann,M., Mosk, A. P., Skoric, B., Pinkse, P. W. H. "Quantum--secure authentication of a physical unclonable key: supplementary material." \textit{Optica} \textbf{1} 421--424 (2014)} i) it is technically infeasible to make a physical clone of a photonic key, ii) an adversary is unable to apply arbitrary unitary transformations to high--dimensional quantum states with low losses (termed quantum--computational unclonability\endnote{Skoric, B. ``Quantum readout of Physical Unclonable Functions: Remote authentication without trusted readers and authenticated Quantum Key Exchange without initial shared secrets.'' \textit{eprint.iacr.org/2009/369}, (2009)}), and iii) different challenges are allowed to be applied to the photonic key. The protocol utilises a challenge/response protocol\endnote{No author. ``Challenge Response Authentication.'' \textit{Wiipedia} (accessed June 2018) \url{https://en.wikipedia.org/wiki/Challenge\%E2\%80\%93response\_authentication}} (i.e. round--by--round interaction between a client and quantum server). Clients request \textit{photonic challenges} from the quantum servers, and due to the interactive nature of authentication and mining, it is in the best interest of quantum servers to support the authentication of clients. The QSA protocol is unique in that it results in an authenticated quantum channel, whilst circumventing the need for pre--shared secrets. This finds various applications, for example, in quantum key growing protocols\endnote{Ling, A., Wildfeuer, C., Scarani, V. ``A note on quantum safe symmetric key growing.'' \\\textit{arXiv:1707.02766v1} (2017)} which may be used to generate symmetric encryption keys for quantum--safe communication without the need for pre--shared secrets.

\subsection{Protocol overview}The photonic challenge is randomly chosen from a set of pre--determined challenges, selected by the quantum server, and is sent to the client's quantum modem via a quantum communication channel for direct interaction with the client's photonic key. Interaction on the client side transforms the incident photonic challenge into a \textit{photonic response}, which is detected on the client side. In accordance with the QSA protocol, the photonic response pairs uniquely with the photonic challenge, provided the challenge physically interacts with the correct photonic key. To guarantee accurate pairing, photonic keys are pre--characterised in a key enrolment stage, with each key requiring $\approx$50MB of characterisation data for accurate enrolment. Enrolment can be done in two ways: i) A certification authority or manufacturer enrols the key. In this case, a key identifier and a precise characterization of the key are entered into a publicly readable and tamper proof database (i.e the blockchain which the key will be used to access). Or, ii) a certificate authority signs a digital certificate containing the key identifier and characterization data. The certificate is then stored publicly, e.g. through inclusion to the blockchain. When a verifier wants to see if a certain key is authentic, the challenge-response behaviour is checked against the enrolled data, which in the case of quantum readout, does not require trust in the remote key reader. The authentication of the quantum channel is then based on the possession of a physical object, instead of knowledge of a pre--shared secret. 
An authentic detection event on the client side occurs when the response wavefront focusses cleanly onto an optical detector. The QSA protocol in its original form assumes that a client's quantum modem is immune to tampering, since an authentic detection event could be easily spoofed if the detector was illuminated in a way that appears consistent with correct authentication. However, the assumption of tampering immunity may be relaxed, provided that the quantum server is allowed to send \textit{decoy} photonic challenges that are randomly and secretly mixed in with real photonic challenges. In this case, because a client is unable to infer whether an incident challenge is legitimate or a decoy (due to the no--cloning theorem of quantum physics), a dishonest client is left uncertain as to whether to correctly illuminate the detector on any given round. This addition of decoy challenges further protects the network from false authentications. Based on experimental demonstrations in literature, in the case of an honest client with a legitimate photonic key, $\sim$200--500 rounds of key interrogation via the challenge/response protocol is sufficient for successful authentication, with authentication concluding in $\approx$200ms--400ms (allowing provision for decoy photonic challenges), with false authentication rates of 1 in 100 billion. 

\appendix 
\section{Experimental Devices and Loopholes}
\noindent Experimental tests of entanglement witnesses can be technologically demanding. To mitigate the technological demands of these tests, it is common practice to make additional assumptions pertaining to the expected action of experimental devices. Importantly, these assumptions lie \textit{outside} of the theoretical framework applied in developing the mathematical inequalities which witness entanglement. Hence, assumptions of this kind pertain only to the experimental devices employed in the test; they may affect the test outcomes (allowing for false positives) but do not affect the veracity of the mathematical derivation. Devices which act in conflict with such assumptions, resulting in false positives (perhaps due to interference by an adversary), are said to open \textit{experimental loopholes}. The key experimental assumptions, which when subverted, open experimental loopholes, are:

\begin{quote}
\textbf{Fair sampling assumption.\endnote{Pearle, P. M. ``Hidden-variable example based upon data rejection.'' \textit{Phys. Rev. D}. \textbf{2} 14181425 (1970)} :}  Non--unit--efficiency detection devices record events that constitute a fair sample of all possible detection events. The \textit{detection loophole} is opened if compromised detectors record a subset of events that are not representative of the entire set. Invocation of the fair--sampling assumption can be avoided by employing high--efficiency detectors, overcoming channel/transmission losses with low--loss channels, or implementing loss--tolerant measurement strategies. 

\textbf{No--signalling assumption:\endnote{Bell, J. S. ``Speakable and Unspeakable in Quantum Mechanics.'' Cambridge, \textit{Cambridge University Press} (1987)}} Experimental devices cannot signal information to each other. The \textit{signalling loophole} is opened if a compromised apparatus communicates secret information to external devices. Invocation of the no--signalling assumption may be avoided by ensuring space--like separation of devices, or by proper experimental design and shielding of devices (a common requirement for cryptographic protocols).

\textbf{Measurement--independence assumption:} Measurement devices and their measurement settings are not pre--programmed or influenced by external sources or events (past or future). The \textit{setting--independence loophole} is opened if devices are externally influenced or their behaviour becomes predictable. Invocation of the measurement--independence assumption may be avoided by ensuring measurement devices employ genuinely random sources of measurement setting, typically enacted with a quantum random number generator (QRNG). \endnote{Gallicchio, J., Friedman, A. S., Kaiser, D. I. ``Testing Bell's inequality with cosmic photons: closing the setting--independence loophole.'' \textit{Phys. Rev. Lett.} \textbf{112} 110405 (2014)}
\end{quote}

\noindent Ensuring closure of the detection, signalling, and freedom--of--choice loopholes is a routine requirement in quantum cryptographic protocols and in applied technologies which utilise tests of quantum correlations.

\theendnotes






\thispagestyle{pagelast}

\end{document}